# Emergent impervious band crossing in the bulk of the topological nodal-line semimetal ZrAs$_2$

A. S. Wadge,[1,*] K. Zberecki,[2,†] B. J. Kowalski,[3] D. Jastrzebski,[1,3,4] P. K. Tanwar,[1] P. Iwanowski,[3]
R. Diduszko,[3] A. Moosarikandy,[1] M. Rosmus,[5] N. Olszowska,[5] and A. Wisniewski[1,3]

[1]*International Research Centre MagTop, Institute of Physics, Polish Academy of Sciences, Aleja Lotnikow 32/46, PL-02668 Warsaw, Poland*
[2]*Faculty of Physics, Warsaw University of Technology, Koszykowa 75, Warsaw, 00-662, Poland*
[3]*Institute of Physics, Polish Academy of Sciences, Aleja Lotnikow 32/46, PL-02668 Warsaw, Poland*
[4]*Faculty of Chemistry, Warsaw University of Technology, Noakowskiego 3, 00-664 Warsaw, Poland*
[5]*National Synchrotron Radiation Centre SOLARIS, Jagiellonian University, Czerwone Maki 98, PL-30392 Cracow, Poland*

Topological nodal-line semimetals represent a unique class of materials with intriguing electronic structures and rich of symmetries, hosting electronic states with nontrivial topological properties. Among these, ZrAs$_2$ stands out, characterized by its nodal lines in a momentum space, governed by nonsymmorphic symmetries. This study integrates angle-resolved photoemission spectroscopy (ARPES) with density functional theory (DFT) calculations to explore the electronic states of ZrAs$_2$. Our study provides experimental evidence of nonsymmorphic symmetry-protected band crossing and nodal lines in ZrAs$_2$. In ARPES scans, we observed a distinctive surface and bulk states at different photon energies associated with nodal lines. Our results, supported by calculations based on DFT, unveil such impervious band crossing anchored at specific points in the Brillouin zone, with particular emphasis on the $S$ point. Surface bands and bulk states near the crossing are elucidated through slab calculations, corroborating experimental findings. These findings enhance our understanding of the electronic structure of ZrAs$_2$.

## I. INTRODUCTION

Topological semimetals (TSMs) represent a novel class of materials in condensed matter physics, distinguished by their unique electronic structures and the intricate role of symmetries. These materials are known for hosting electronic states that exhibit nontrivial topological properties [1–5]. In their electronic structure, the bands meet at certain isolated points or lines in the momentum space, giving rise to anomalous contribution to the physical properties, such as Shubnikov–de Haas oscillations, ultrahigh mobility, extremely large magnetoresistance, large anomalous Hall conductivity, etc. [6,7]. Angle-resolved photoemission spectroscopy (ARPES) plays a pivotal role in this research field by providing direct insight into the electronic band structures of the materials, enabling us to observe bulk and surface states and characterize symmetry- protected states [8–11].

Nodal-line semimetal is one of the types of TSMs where the conduction and valence bands intersect not at discrete points, but along a continuous loop in the momentum space, forming a one-dimensional "nodal line" (NL). NLs mark the presence of gapless bands, where two separate bands intersect accidently. The topological properties of accidental line nodes have an even codimension and lack absolute topological stability. However, by imposing certain symmetries, these line nodes can be stabilized, even if they are not necessarily positioned at the Fermi energy [12]. Hence observing the NL poses a challenging task for experiments. Nonsymmorphic symmetry, which involves a partial lattice shift combined with either a reflection (glide plane) or a rotation (screw axis), is sparsely seen in semimetals. The presence of nonsymmorphic symmetry only ensures the highly degenerate band crossings in the material which are resilient to spin-orbit interaction. The necessary condition to fix these crossings either at Γ point or at any high symmetry point is that material should possess nonsymmorphic symmetry as well as inversion symmetry [13–16].

Burkov *et al.* (2011) [12] presented a theoretical investigation of "nodal-semimetal" phases, specifically focusing on Weyl semimetals and line-node semimetals. Kim *et al.* (2015) [13] introduced a $Z_2$ class of topological semimetals characterized by Dirac line-nodes (DLNs) in inversion-symmetric crystals. Fang *et al.* (2015) [14] studied two classes (with and without SOC) of three-dimensional TSMs with nodal lines protected by crystalline symmetries. Their theoretical contribution broadened the understanding of the role of crystalline symmetries in stabilizing nodal lines and suggested the potential for experiments [17–23].

In 2016, Schoop *et al.* [24] conducted a comprehensive theory and ARPES study on ZrSiS, revealing a Dirac cone protected by nonsymmorphic symmetry and the presence of three-dimensional Dirac lines. In the same year, Takane *et al.* [25] unearthed a Dirac node arc in the topological nodal-line semimetal HfSiS, a finding that emphasized the intricate nature and promising capabilities of HfSiS. Further advancements by Sims *et al.* [26] revealed surface states in HfP$_2$ which vary depending on the material's termination. Bannies *et al.* [27] enhanced our knowledge through ARPES spectra studies on $ZrP_2$ in which bulk Fermi surface was observed on (001) orientation. Simultaneously, in the same year, Hao *et al.* [28] identified multiple Dirac nodal lines in $TaNiTe_5$, further broadening our perspective. Wu *et al.* [29] identified nonsymmorphic symmetry-protected band crossings in the square net metal $PtPb_4$. Mohanta *et al.* [30] and Fakherdine *et al.* [31] theoretically showed the nonsymmorphic symmetry protected semi-Dirac crossings fixed at the boundary of the Brillouin zone.

---

*Contact author: wadge@magtop.ifpan.edu.pl
†Contact author: krzysztof.zberecki@pw.edu.pl

Zhou *et al.* [8] provided theoretical insights into nonsymmorphic symmetry and unprecedented butterflylike nodal lines with topological nature on the (010) plane of $ZrAs_2$. Additionally, our preliminary transport studies (Wadge *et al.* [32]) demonstrated pronounced Shubnikov–de Haas (SdH) oscillations above 9 T. The fast Fourier transform (FFT) of these SdH oscillations indicated the presence of two distinct bands at 195 and 682 T, suggesting the existence of two electronic pockets. Interestingly, a very recent study on magnetotransport has reported a significant SdH oscillation peak at 567 T, further underscoring the complexity and richness of the electronic structure of $ZrAs_2$ [33].

This study motivated an investigation into the band structure of $ZrAs_2$ using ARPES. Due to the natural cleavage of $ZrAs_2$ at the (001) plane, we focused our study on the band structure and corresponding ARPES for the (001) plane. This study reports the investigation of the electronic structure of single crystals of $ZrAs_2$. We demonstrated the Fermi surface using ARPES and DFT calculations. Additionally, we aimed to study nonsymmorphic symmetry-protected band crossing anchored at the $S$ point in the band structure.

## II. EXPERIMENTAL DETAILS

### A. Single crystal growth and structural characterization

Single crystals of $ZrAs_2$ were grown using the two-stage chemical vapor transport (CVT) method, as shown in Fig. 1. Initially, polycrystalline $ZrAs_2$ was synthesized through the direct reaction of Zr sponge (Koch-Light Laboratories Ltd, 99.8%) and As (PPM Pure Metals, 99.999 995%) in an evacuated quartz ampoule. The ampoule was maintained for 7 days the internal pressure. The resulting polycrystalline $ZrAs_2$ was pressed into pellets, loaded into a quartz tube with iodine (POCH, 99.8%, 10 $mg/cm^3$ of ampoule volume) as transport agent, and sealed under vacuum. This assembly was then placed in a furnace with a temperature gradient of 700 °C (source zone) and 800 °C (crystallization zone) for 21 days. Afterward, the furnace was turned off and allowed to cool down to room temperature. Crystal structure and crystallographic quality of grown $ZrAs_2$ were verified by x-ray powder diffraction using a Rigaku SmartLab 3kW diffractometer equipped with a tube having Cu anode, and operating with $U = 40$ kV and $I = 30$ mA. The characteristic peak positions were identified, as shown in Fig. 1(c), based on the data from International Centre for Diffraction Data (ICDD), Powder Diffraction File PDF-4+2023 RDB database.

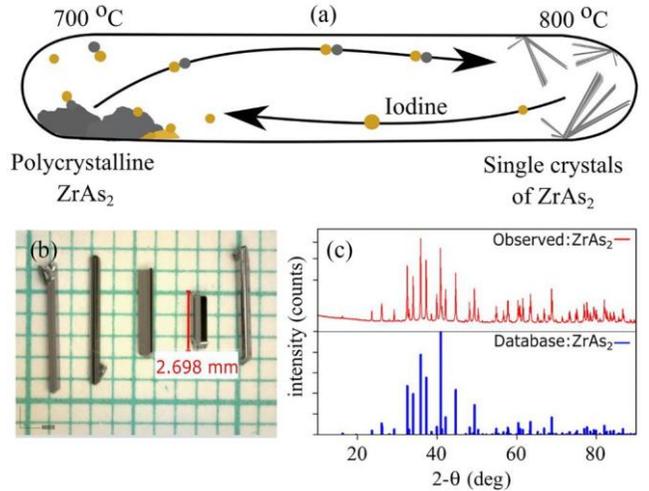

FIG. 1. (a) Schematic of the chemical vapor transport for $ZrAs_2$ synthesis, with source zone at 700 °C and growth zone at 800 °C, (b) needlelike $ZrAs_2$ single crystals obtained through the process, and (c) powder x-ray diffraction of the crystals.

For the quantitative chemical analysis of $ZrAs_2$, energy dispersive x-ray spectroscopy (EDX) was employed using a QUANTAX 400 Bruker system coupled with a Zeiss Auriga field-emission (Schottky-type) scanning electron microscope. The EDX measurements were conducted at 15 kV incident energy. The sample was prepared by sticking the crystals of $ZrAs_2$ on the carbon tape. Oriented samples were utilized, and the results confirmed the stoichiometric composition of $ZrAs_2$, with an atomic ratio of 1:2 (see Supplemental Material Fig. 1) [34], within the experimental error margin.

### B. ARPES setup

Our experiment was conducted using the URANOS beamline at the National Synchrotron Radiation Centre SOLARIS in Krakow, Poland. The radiation source for our measurements was a quasiperiodic elliptically polarizing APPLE II type undulator, emitting photons in the energy range 8–170 eV. The facility was equipped with a SCIENTA OMICRON DA30L photoelectron spectrometer. This spectrometer has a maximum detector resolution and angular resolution of 1.8 meV. However, the practical resolution depends on several factors such as the exit slit, photon energy, CFF, and pass energy. For this experiment, we achieved an energy resolution of 17 meV at 30 eV photon energy and 67 meV for 100 eV photon energy. The spectrometer also has an angular resolution of 0.1° and features deflectors that allow for wide-angle band structure measurements without repositioning the sample. This setup enabled us to perform precise band mapping across the entire Brillouin zone (BZ). ARPES spectra were obtained from single crystals that were freshly cleaved *in situ* within an ultrahigh vacuum (UHV) ($\sim 3.7 \times 10^{-11}$ torr) environment to obtain (001) surface, using different photon energies ranging 30–100 eV. During these measurements, the samples were consistently maintained at a temperature of

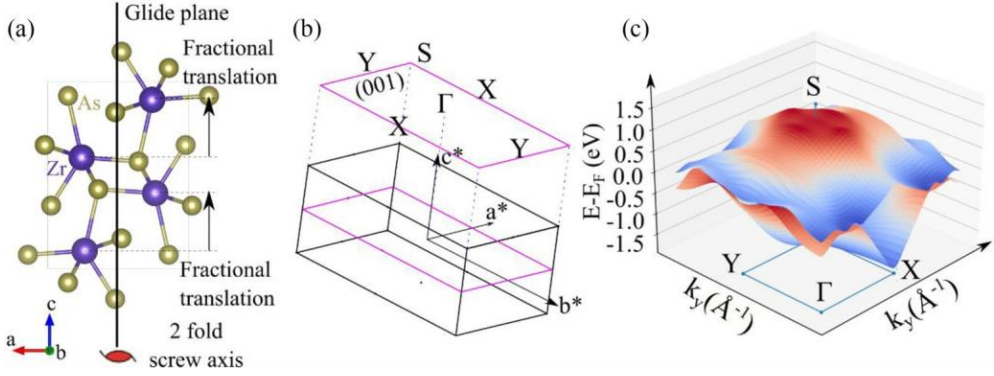

FIG. 2. (a) Crystal structure of ZrAs$_2$ showing nonsymmorphic symmetry with glide plane and twofold screw axis, (b) 3D Brillouin zone and projection on (001) plane with high symmetry points, (c) 3D visualization of the bands with SOC.

80 K. The analysis of the gathered ARPES data was conducted using IGOR PRO software and Python scripts.

### III. DFT CALCULATION DETAILS

All the calculations were performed within density functional theory (DFT) as implemented in the VASP package [35–38] with projector augmented wave pseudopotential (PAW) [39,40], Perdew-Burke-Ernzerhof (PBE), and generalized gradient approximation (GGA) functionals [41]. For the sampling of the Brillouin zone, a dense 8 × 8 × 8 grid was used, while the plane wave energy cutoff was set to 520 eV. All the structures were optimized until the force exerted on each ion was smaller than $10^{-5}$ eV/Å. Three-dimensional (3D) plots of Fermi surface were made with use of the XCrysDen code [42]. For further evaluation of the DFT results, the tight binding (TB) model was prepared with use of the WANNIER90 code [43]. This model was used to calculate surface states of the slab structure with use of the Wannier Tools package [44]. For the surface termination calculations, a 1 × 1 × 4 supercell was employed, incorporating a 15 Å vacuum layer on top to model the surface. The total energy calculations indicated that "term. 2" was the most stable configuration.

### IV. RESULTS AND DISCUSSION

The CVT process produced needlelike single crystals of ZrAs$_2$ of average length 5–6 mm as shown in Fig. 1(b). X-ray diffraction analysis showed that ZrAs$_2$ crystallizes in an orthorhombic crystal structure with space group number 62 (*pnma*, $D_{2h}$). It is centrosymmetric in nature with the refined lattice parameters $a = 6.801\,63(10)$ Å, $b = 3.688\,91(5)$ Å, $c = 9.030\,58(13)$ Å. In crystal structure of ZrAs$_2$, the Zr ion (blue) is surrounded by nine As ions (tan). It has nonsymmorphic symmetry with $n$ glide plane perpendicular to "$a$" axis and twofold rotation screw axis [see Fig. 2(a)] [8]. The x-ray diffraction data confirm the single crystalline phase and are consistent with previous studies [33,45]. The lattice constants obtained through relaxation processes in computational simulations align closely with the values predicted by density functional theory (DFT). In Fig. 2(b), we present the (001) surface projection of the three-dimensional Brillouin zone, highlighting high symmetry points with $\Gamma$ at the center. Figure 2(c) depicts a three-dimensional visualization of the electronic band structure, illustrating the dispersion of energy states across $k_x$ and $k_y$ momentum space.

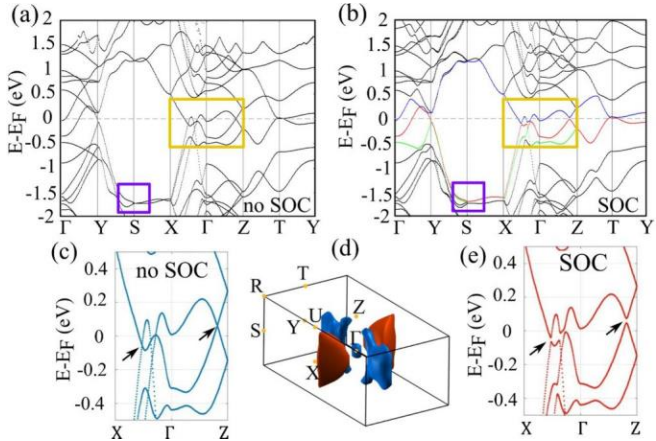

FIG. 3. Bulk band structures of ZrAs$_2$ calculated by density functional theory (DFT). (a) Band structure without spin-orbit coupling (SOC), showing accidental band crossings (yellow) and nonsymmorphic symmetry protected band crossings (purple). (b) Band structure with SOC. (c) Magnified view of the band structure along the X-Γ-Z path without SOC, highlighting the band crossings. (d) 3D Brillouin zone of ZrAs$_2$ showing electron pockets (blue) and hole pockets (red). (e) Magnified view along the X-Γ-Z path with SOC, demonstrating the gap opening due to SOC by arrows.

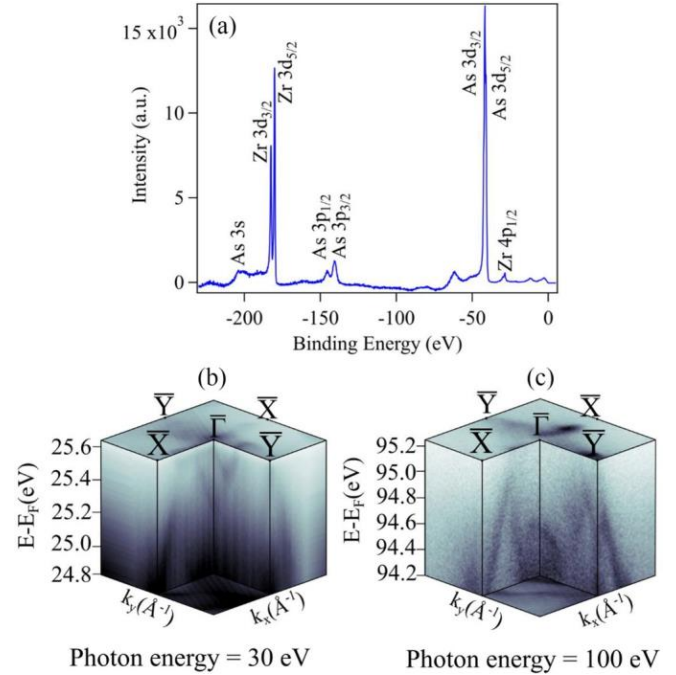

FIG. 4. (a) Core-level spectra of ZrAs$_2$ indicating Zr and As peaks confirming the only elements in ZrAs$_2$. 3D visualization of ARPES data for ZrAs$_2$ obtained at photon energies: (b) 30 eV and (c) 100 eV.

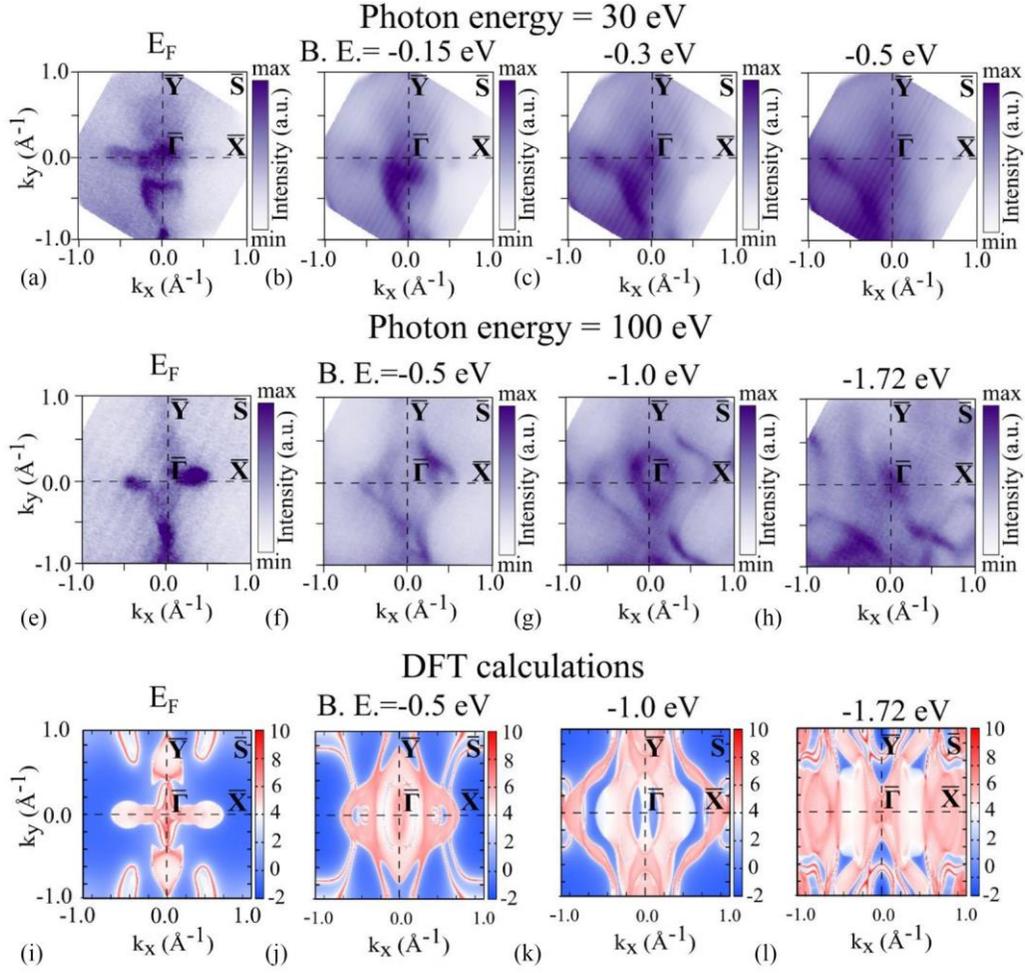

FIG. 5. The constant energy contours of ZrAs$_2$ at different binding energies obtained using ARPES. (a)–(d) Contours at 30 eV photon energy for binding energies: (a) $E_F$, (b) —0.15 eV, (c) —0.3 eV, and (d) —0.5 eV. (e)–(h) Comparison of constant energy contours at 100 eV photon energy with corresponding DFT-calculated constant energy contours for binding energies: (e), (i) $E_F$, (f), (j) —0.5 eV, (g), (k) —1.0 eV, and (h), (l) —1.72 eV.

DFT band structure calculations were performed both without and with the inclusion of SOC, as depicted in Figs. 3(a) and 3(b). The figures illustrate two distinct types of band crossings: accidental band crossings (ABCs), highlighted by a yellow rectangle, and nonsymmorphic symmetry-protected band crossings, highlighted by a purple rectangle. In the absence of SOC, band crossings can be observed along the X-$\Gamma$-Z path [see magnified bands for better visibility in Figs. 3(c) and 3(e)]. These crossings are disrupted and a gap of ~20 meV emerges upon the inclusion of SOC. Conversely, the band crossings remain unaffected by SOC, as they are safeguarded from opening the gap by nonsymmorphic symmetry. Figure 3(d) illustrates the 3D Brillouin zone with electron (blue) and hole (red) Fermi pockets. Unlike ZrP$_2$, ZrAs$_2$ lacks a third pocket associated with green band (hole-type) with the lowest energy [27]. Though ABC opens up the gap, which is very small, it preserves the semimetallic nature as described in Nandi *et al.* [33]. The bands near the Fermi level along the $\Gamma$-X path comprise a mix of As ($p$) and Zr ($d$) states, preventing assignment to specific ions (see Supplemental Material Fig. 8) [34].

The core-level spectrum was taken prior to ARPES measurements in order to check the composition and they showed the significant peaks of Zr $3d_{3/2}$ (181.1 eV), Zr $3d_{5/2}$ (178.8 eV), Zr $4p_{1/2}$ (28.5 eV), as well as As $3s$ (204.7 eV), As $3p_{1/2}$ (146.2 eV), As $3p_{3/2}$ (141.2 eV), As $3d_{3/2}$ (41.7 eV), and As $3p_{5/2}$ (41.7 eV), which are in good agreement [see Fig. 4(a)] with the binding energy database [45]. These peaks confirm the composition of Zr and As with no foreign

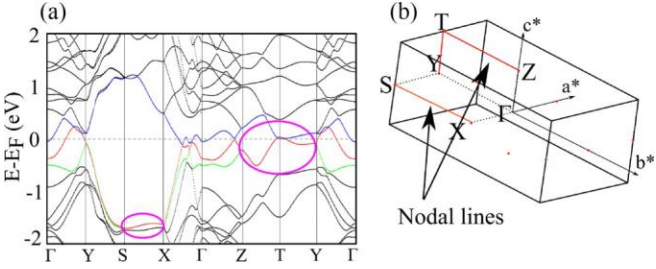

FIG. 6. (a) The bulk band structure in which nodal lines are indicated by magenta color ellipses, (b) the paths of nodal lines are shown in the Brillouin zone.

quantity in the crystals.

We obtained ARPES spectra at various photon energies, with the most significant results observed at 30 and 100 eV [see Figs. 4(b) and 4(c)]. At higher photon energies, specifically 100 eV, we were able to map a more extensive portion of the Fermi surface, including the $S$ points. In contrast, measurements at 30 eV revealed a smaller region of the Brillouin zone, which precluded the observation of the S point. Later, it turned out that $ZrAs_2$ naturally cleaves at the (001) plane with six possible terminations (Supplemental Material Figs. 5 and 6) [34].

Figures 5(a)–5(d) show the constant energy contours at a photon energy of 30 eV, illustrating the evolution from the Fermi level to a binding energy of $E_B = -1.06$ eV at various binding energies. We have obtained ARPES at photon energies ranging 34—54 eV. As the photon energy changes, we observed shifts in the positions of several bands, indicating that these are likely bulk bands. In contrast, some bands at a binding energy of —0.5 eV remain stationary, suggesting they are surface states considering the overall spectra; the electronic structure is predominantly bulk in nature as shown in Supplemental Material Fig. 4 [34].

In addition, Figs. 5(e)–5(l) compare the constant energy contours at 100 eV photon energy with corresponding DFT-calculated contours for binding energies: (e) and (i) $E_F$, (f) and (j) —0.5 eV, (g) and (k) —1.0 eV, and (h) and (l) —1.72 eV. The theoretical constant energy contours were obtained for the semi-infinite slab with orientation (001). For $E = E_F$, the electron and hole pockets [see Fig. 3(b)] can be clearly visible. Lowering the energy, the hole pocket shrinks, while the electron pocket becomes more robust. For $E_B = -1.72$ eV in the vicinity of the $S$ point one can observe the pronounced signatures of the Dirac crossings.

Figure 6 presents DFT analysis of the bulk band structure of $ZrAs_2$ and the characterization of its nodal lines. In Fig. 6(a), the bulk band structure is shown, with nodal lines highlighted by magenta ellipses along the X-S and Z-T-Y paths. Additionally, nodal lines are indicated in red within the 3D Brillouin zone. The path $\overline{Z} - \overline{T} - \overline{Y}$ is not observable in the ARPES results because the high-symmetry points Z and T are not present on the (001) plane. However, the nodal line along the $\overline{S} - \overline{X}$ path is clearly visible in Supplemental Material Figs. 3(a) and 3(b) [34].

Figures 7(a) and 7(e) shows the Fermi surface at photon energy 30 and 100 eV with all the high symmetry points. The cut along $\overline{X} - \overline{\Gamma} - \overline{X}$ reveals the band dispersion near the $\overline{\Gamma}$ point [see Figs. 7(b) and 7(c)], and the cut along $\overline{S} - \overline{Y} - \overline{S}$ is also shown. The consistency between these cross-sectional cuts and the slab calculations, displayed in Figs. 7(b) and 5(f) respectively, is notable.

Another important finding of our work is nonsymmorphic symmetry protected band crossings in the bulk. Nonsymmorphic symmetry alone ensures only the existence of band crossing in the band dispersion, yet it does not determine the specific location of this degenerate point in momentum space. However, the presence of inversion symmetry anchored the band crossing either at the $\bar{\Gamma}$ point or along the boundary of the Brillouin zone (BZ) [14]. Along with time reversal symmetry and inversion symmetry, ZrAs$_2$ shows nonsymmorphic symmetry. It has reflection with partial translation, $G = \{R|\vec{t}\}$, where G is glide symmetry, R is reflection and t is nonprimitive translation, and $S = \{C_2|\vec{t}\}$, where S is screw symmetry, and $C_2$ is twofold rotation.

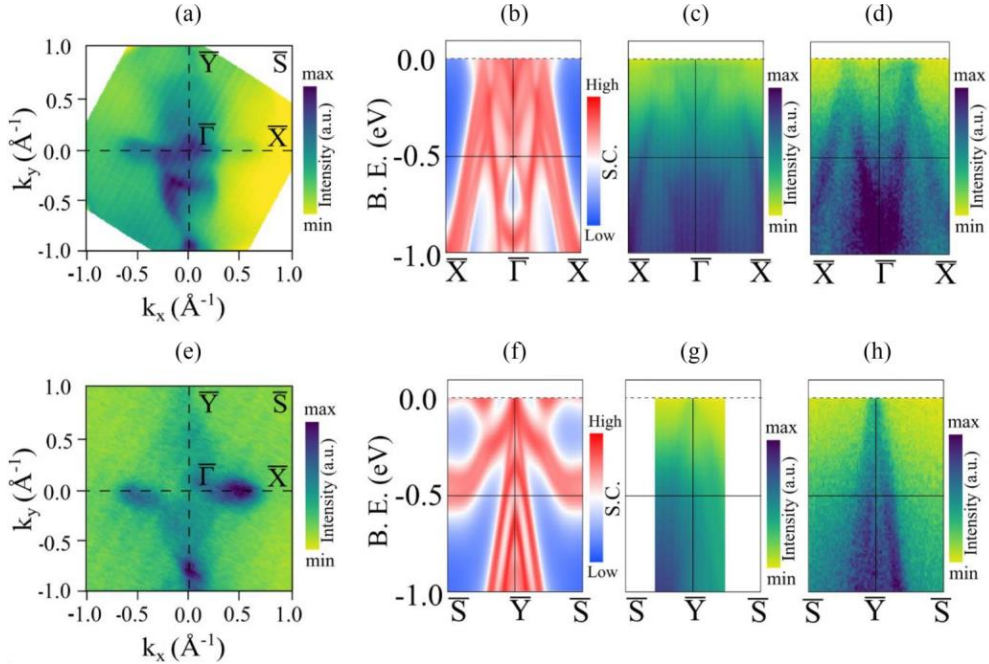

FIG. 7. (a), (e) ARPES reveals 2D Fermi surfaces at 30 and 100 eV photon energies. (b), (f) Theoretical band dispersion along $\overline{X} - \overline{\Gamma} - \overline{X}$ and $\overline{S} - \overline{Y} - \overline{S}$ paths. ARPES cuts at (c), (d) 30 eV and (g), (h) 100 eV photon energies, respectively.

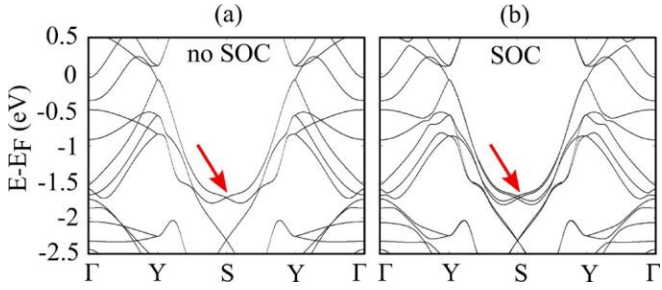

FIG. 8. Bulk band dispersion (a) without SOC and (b) with SOC along Γ-Y-S-Y-Γ showing Dirac band crossing (red arrow) at S point (−1.72 eV).

The DFT-calculated bulk band dispersions, shown in Figs. 8(a) and 8(b), reveal that the band crossing at the $\bar{S}$ point exhibits fourfold degeneracy in the absence of spin-orbit coupling (SOC). Due to the protection of nonsymmorphic symmetry against SOC, there are twofold degenerate crossings at the $\bar{S}$ point ($E_B = -1.72$ eV), as illustrated in Fig. 8. The ARPES results presented in Figs. 9(a) and 9(b) confirm the presence of such band crossings at the $\bar{S}$ point. Additionally, various perpendicular cuts systematically demonstrate that at the $\bar{S}$ point, band crossings occur, and as the measurements move away from the $\bar{S}$ point, a gap begins to open. These observations are consistent with the DFT bulk and slab calculations shown in Figs. 9(c) and 9(d), providing a comprehensive understanding of the electronic structure. According to Zhou et al. [8], ZrAs$_2$ hosts topologically protected nodal lines consisting of concentric intersecting coplanar ellipses on the (010) plane whereas Nandi et al. [33] showed the nontrivial Berry phase from the transport measurements also on the (010) plane. Demonstrating the nontriviality of the bands on the naturally cleaved (001) plane is challenging because most of the surface and bulk bands overlap.

Although we have not yet obtained ARPES spectra on other cleavage planes, we have calculated the Fermi surfaces for the (100), (010), and (001) planes (see Supplemental Material Fig. 7) [34]. This opens up the possibility for more detailed ARPES studies on cleavage planes other than (001).

## V. CONCLUSIONS

In this study, we comprehensively analyzed the electronic structure of ZrAs$_2$ using DFT calculations and ARPES measurements. Our observations revealed a 3D band structure characterized by electron and hole pockets, significantly influencing the material's electronic properties. The DFT calculations, including spin-orbit coupling, identified both accidental band crossings (ABCs) and nonsymmorphic symmetry protected band crossing.

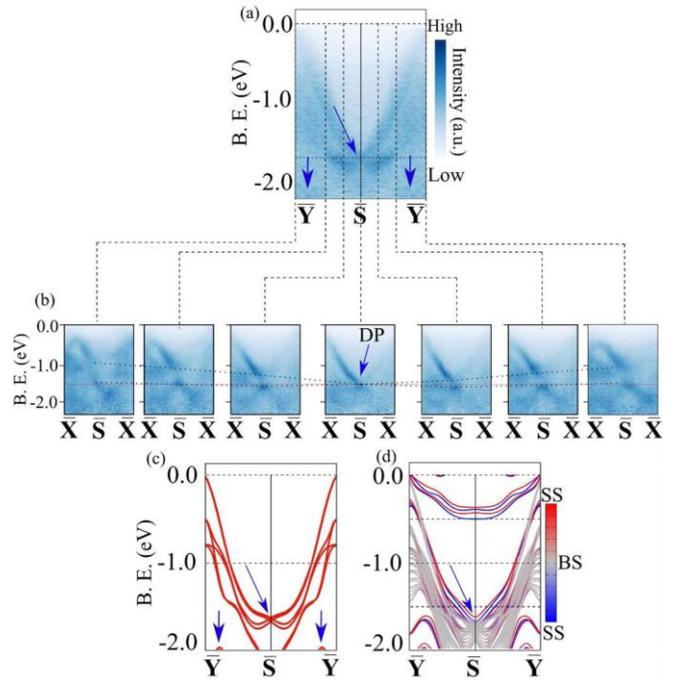

FIG. 9. (a) Illustrating potential Dirac-like crossings at the $\bar{S}$ points along the $\bar{Y}-\bar{S}-\bar{Y}$ path. (b) Multiple slices along the $\bar{X}-\bar{S}-\bar{X}$ direction, demonstrating band evolution across the $\bar{Y}-\bar{S}-\bar{Y}$ vector. (c), (d) Bulk band calculations and slab calculations along the $\bar{Y}-\bar{S}-\bar{Y}$ path (SS: surface states and BS: bulk states).

The ARPES data, obtained from *in situ* cleaving at the (001) plane, confirmed the presence of nodal lines along the $\bar{S}-\bar{X}$ path. Notably, we observed a nonsymmorphic symmetry-protected band crossing at $E_B = -1.72$ eV at the $\bar{S}$ point, indicating dominant bulk states. However, due to the cleaving at the (001) plane, we could not observe other parts of the nodal line along the $\bar{Z}-\bar{T}-\bar{Y}$ path.

This comprehensive investigation sheds light on the intricate electronic behavior of ZrAs$_2$ with the involved symmetries, important for fundamental understanding of nodal-line semimetals especially, and the critical role of nonsymmorphic symmetry in protecting band crossings at the S point. These insights could inform future research and potential applications in topological quantum materials and devices.

### ACKNOWLEDGMENTS

We are grateful to B. Turowski for his invaluable assistance with ARPES cube data procedures, and to K. Cieslak for her support at the beamline. We thank C. Autieri, R. Bacewicz, C. Jatrzebski, and W. Paszkowicz for their insightful discussions. This research was supported by the Foundation for Polish Science project "MagTop" (Project No. FENG.02.01- IP.05–0028/23) cofinanced by the European Union from the funds of Priority 2 of the European Funds for a Smart Economy Program 2021–2027 (FENG). We acknowledge with appreciation the SOLARIS Centre for providing access to Beamline UARPES, where the crucial ARPES measurements were conducted. Additionally, we express our appreciation to Poland's high-performance infrastructure PLGrid, operated by Cyfronet AGH, for providing computational facilities and support under Computational Grant No. plgnanotrans10.

# Supplemental Material

# Emergent impervious band crossing in the bulk of the topological nodal-line semimetal ZrAs$_2$

A. S. Wadge[1*], K. Zberecki[2†], B. J. Kowalski[3], D. Jastrzebski[1,3,5], P. K. Tanwar[1], P. Iwanowski[3], R. Diduszko[3], A. Moosarikandy[1], M. Rosmus[4], N. Olszowska[4] and A. Wisniewski[1,3]

[1] *International Research Centre MagTop, Institute of Physics, Polish Academy of Sciences, Aleja Lotnikow 32/46, PL-02668 Warsaw, Poland*
[2] *Faculty of Physics, Warsaw University of Technology, Koszykowa 75, Warsaw, 00-662, Poland*
[3] *Institute of Physics, Polish Academy of Sciences, Aleja Lotnikow 32/46, PL-02668 Warsaw, Poland*
[4] *National Synchrotron Radiation Centre SOLARIS, Jagiellonian University, Czerwone Maki 98, PL-30392 Cracow, Poland*
[5] *Faculty of Chemistry, Warsaw University of Technology, Noakowskiego 3, 00-664 Warsaw, Poland*

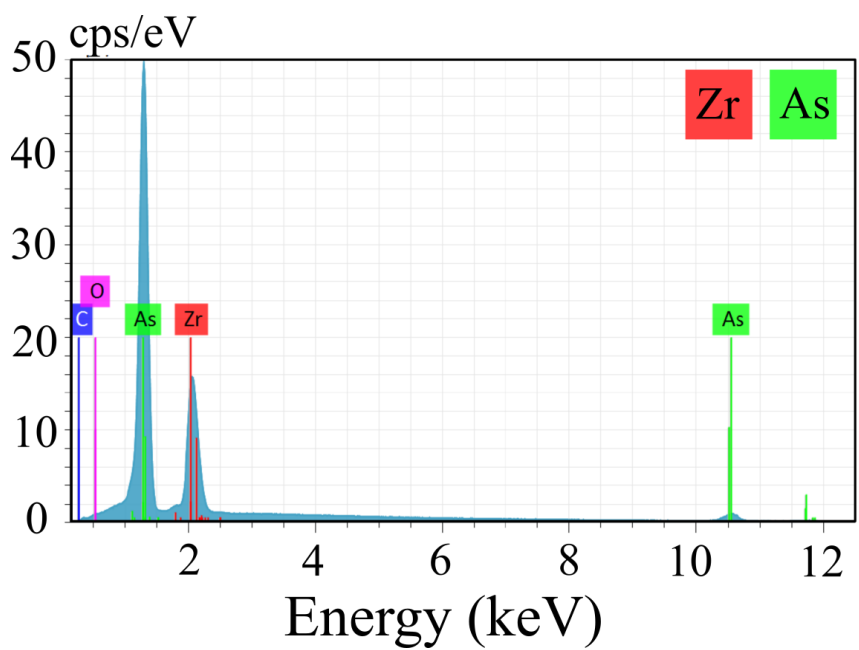

**Suppl. Figure 1** EDX spectrum taken on the shown surface presenting significant Zr and As peaks.

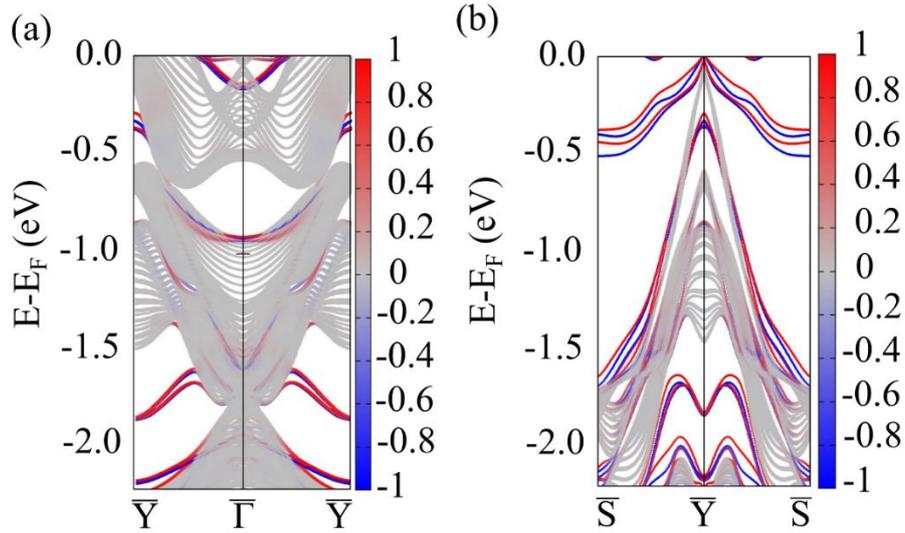

**Suppl. Figure 2** The slab calculations along (a) $\bar{Y} - \bar{\Gamma} - \bar{Y}$ and (b) $\bar{S} - \bar{Y} - \bar{S}$.

**Nodal line:**

On (001) cleaved plane, the nodal line along the $S-X$ path is clearly visible in Suppl. Figure 3 (a). This figure shows ARPES images of the bands, providing experimental evidence of the nodal line along the $S-X$ direction. To further elucidate the nature of these nodal lines, Suppl. Figure 3 (b) demonstrates that the nodal line is a collection of Dirac crossings. This tracking of Dirac points along specific paths in the Brillouin zone helps to observe the nodal line.

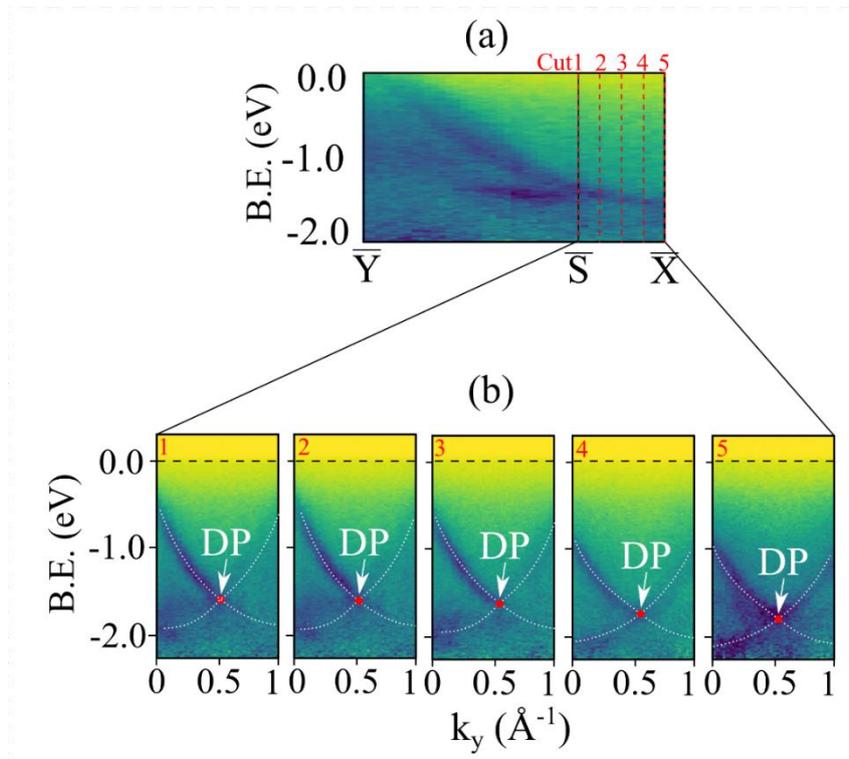

**Suppl. Figure 3** (a) ARPES images of bands with nodal line along $\bar{S} - \bar{X}$ and (b) the nodal line is the collection of Dirac points (DP).

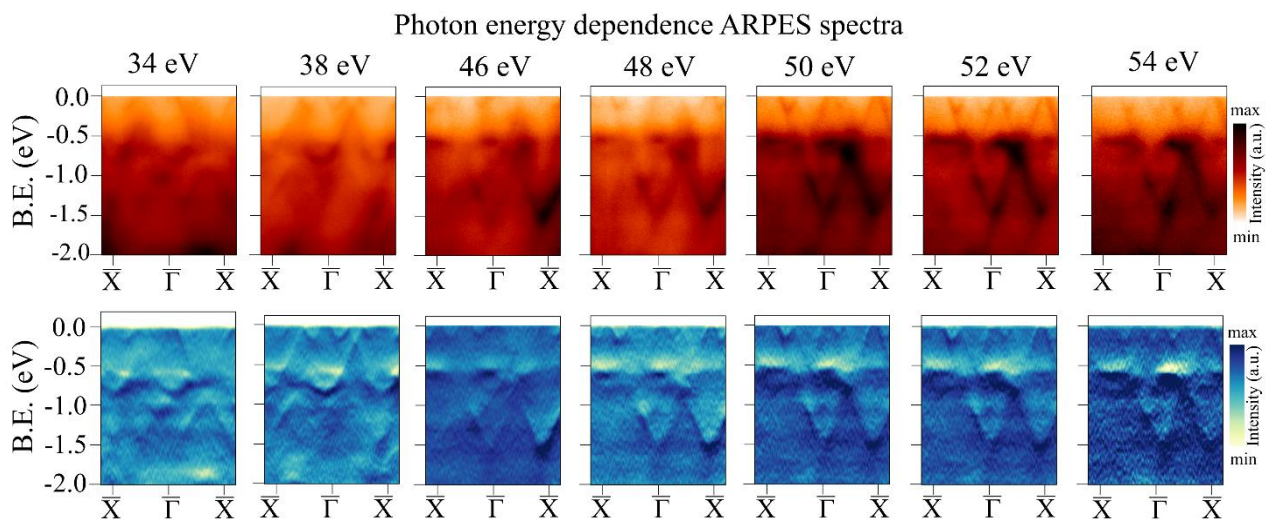

**Suppl. Figure 4** Tracking photon energy variations along the $\overline{X} - \overline{\Gamma} - \overline{X}$ path from 34 $eV$ to 54 $eV$, alongside the associated 2D curvatures.

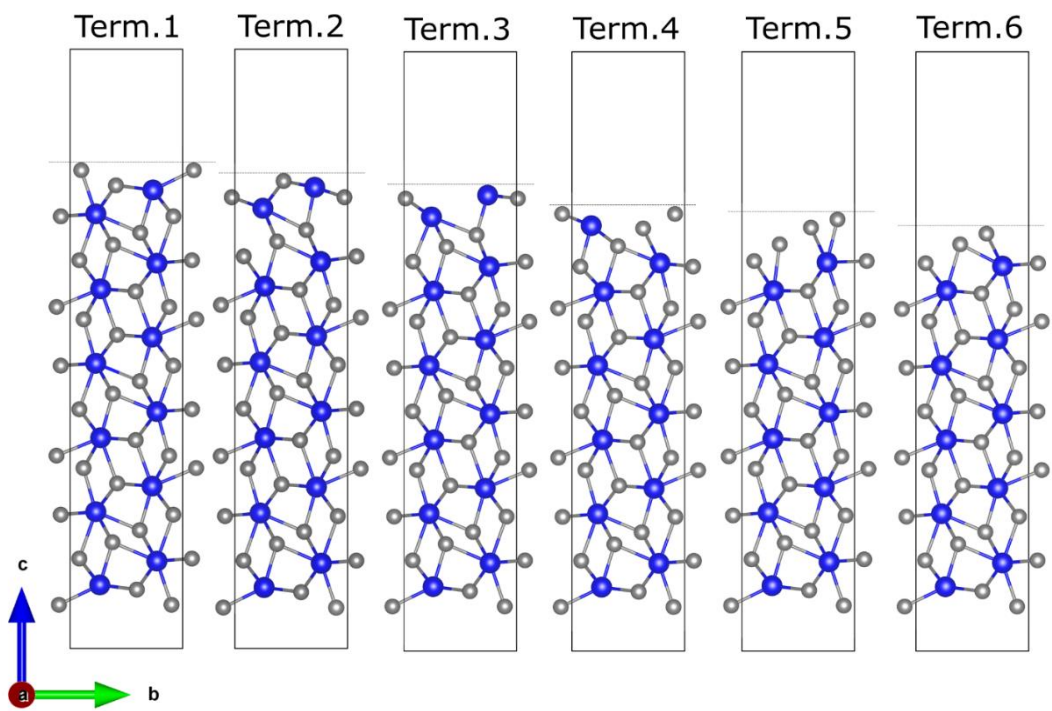

**Suppl. Figure 5** illustration of different terminations (from term. 1-6) perpendicular to the *c* axis

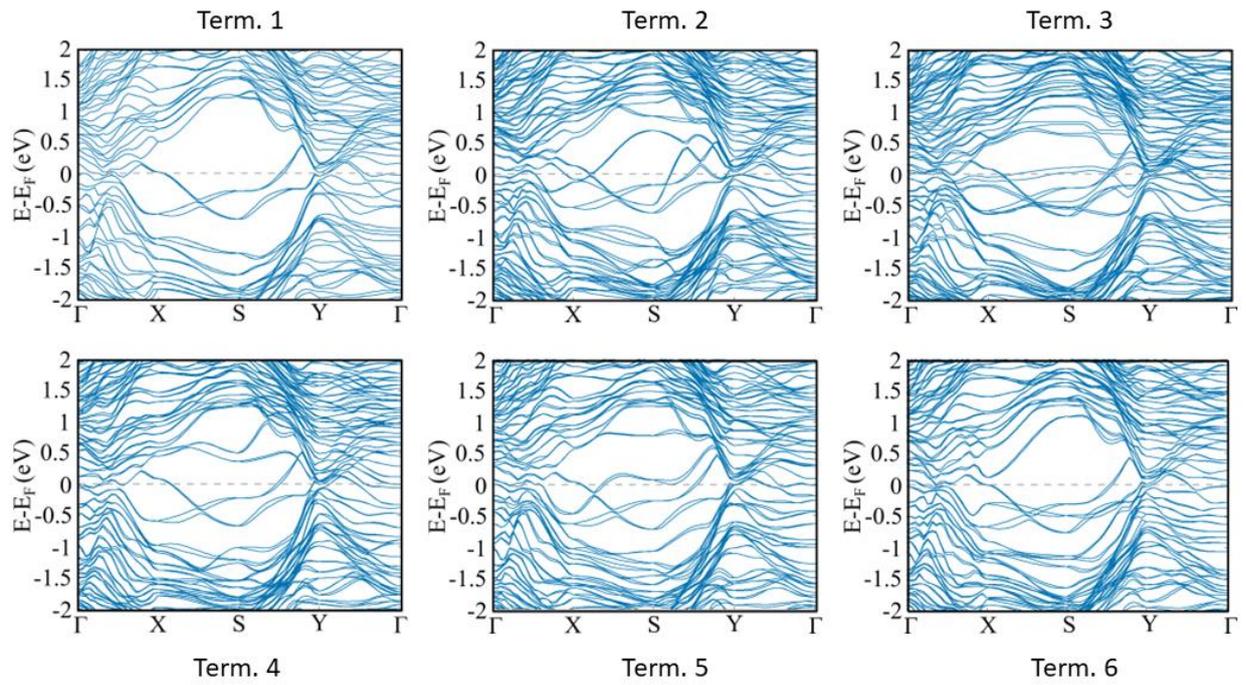

**Suppl. Figure 6** theoretically calculated band structures with surface and bulk contribution for all terminations (term. 1-6) with term.1 as the best possible match with our data.

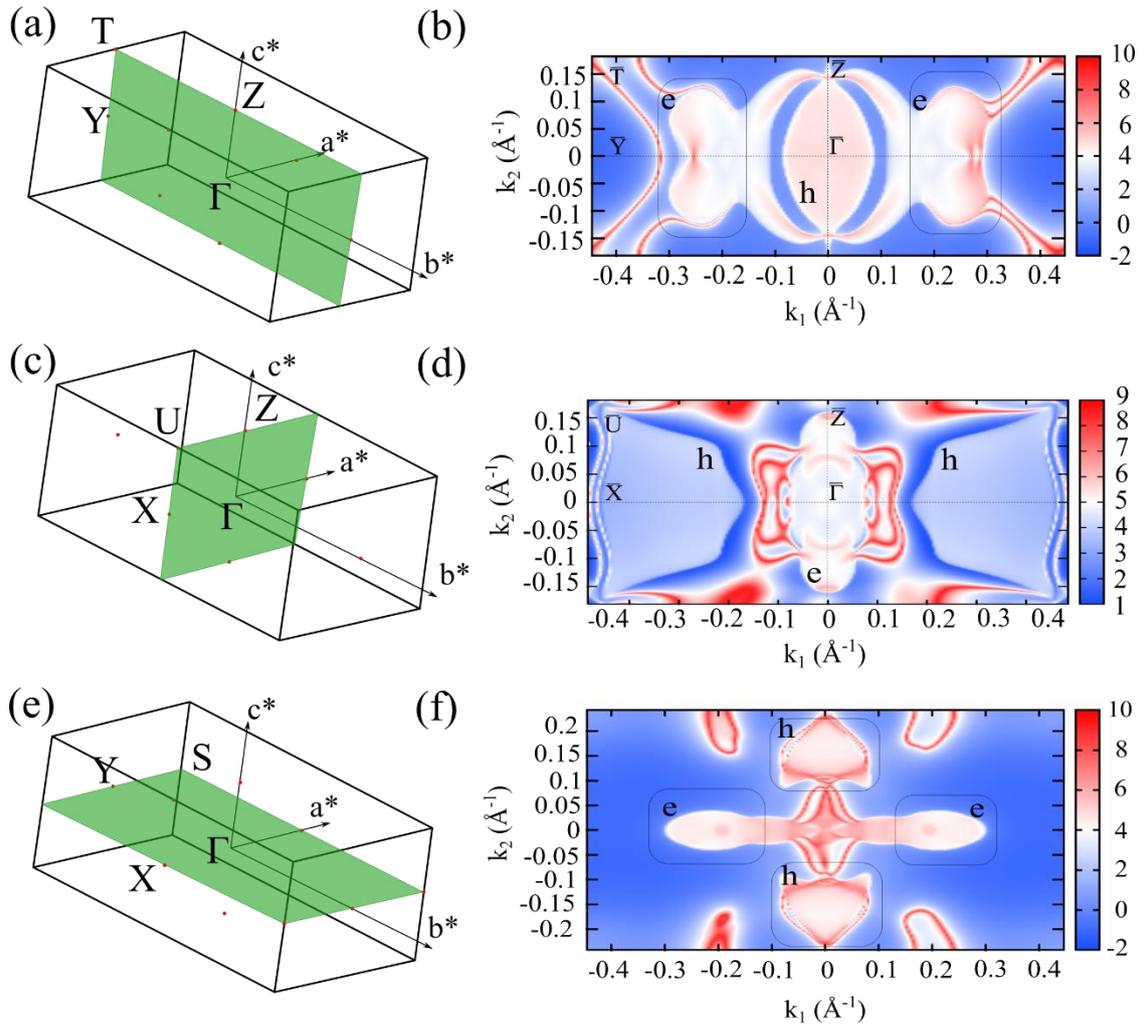

**Suppl. Figure 7** Brillouin zone showing plane along (a) (100), (c) (010), (e) (001) indicated by green color planes with theoretically calculated Fermi surface (E = $E_F$) with $\Gamma$ at the center on (b) (100) and (d) (010) and (f) (001) plane indicating electron 'e' and hole 'h' pockets.

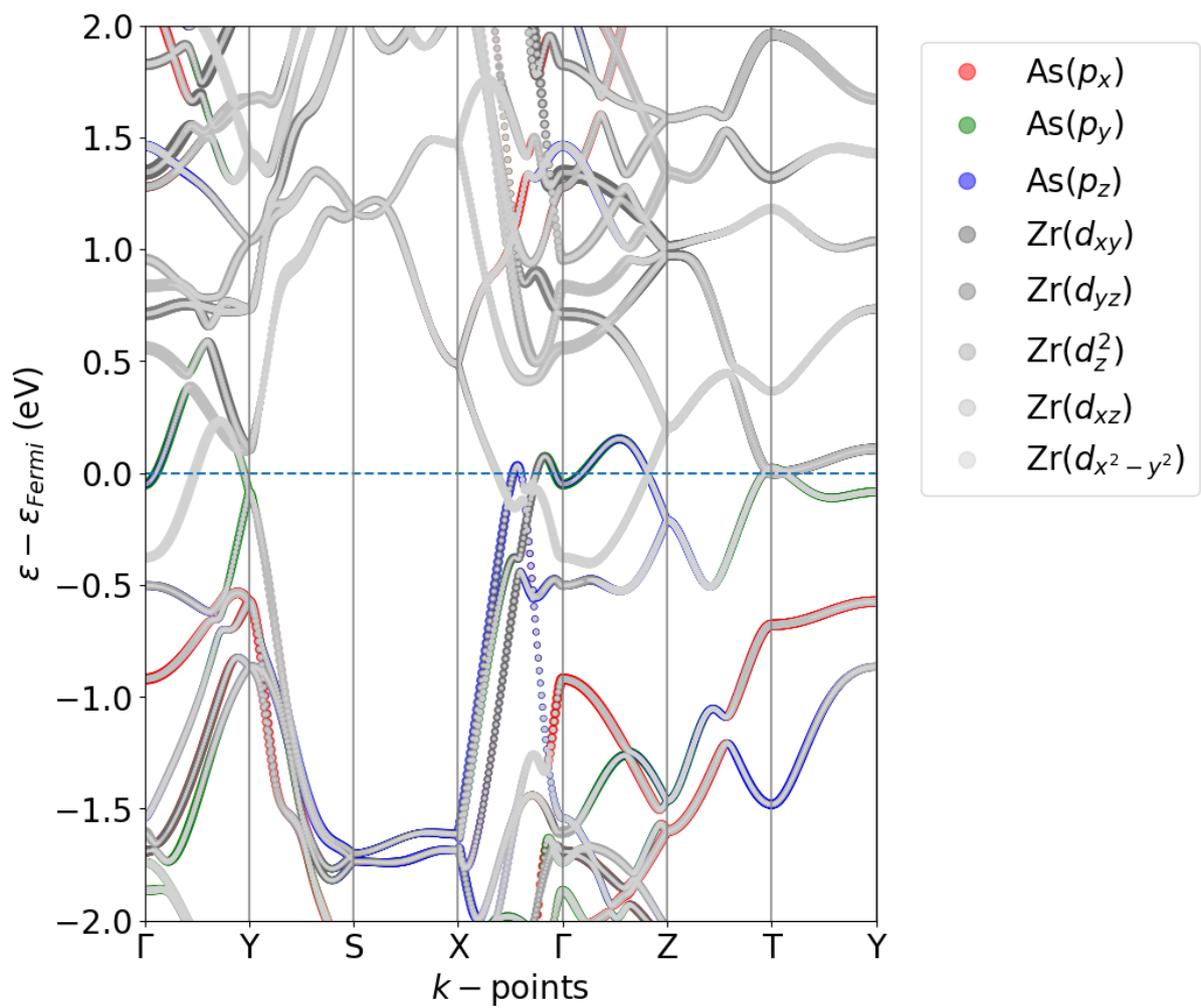

**Suppl. Figure 8** Bulk band structure illustrating Zr ($d$-orbital) and As ($p$-orbital) contributions.